\documentclass[%
 reprint,
 amsmath,amssymb,
 aps,nofootinbib
]{revtex4-2}

\usepackage{graphicx}
\usepackage{dcolumn}
\usepackage{bm}

\usepackage{hyperref}
\hypersetup{colorlinks=true,linkcolor=blue,urlcolor=blue}

\begin{document}

\preprint{APS/123-QED}

\title{The Conservation of Mechanical Energy simulator -- SimuFísica}
\thanks{\url{https://simufisica.com/en/conservation-mechanical-energy/}}%

\author{Marco P. M. de Souza}
\email{marcopolo@unir.br}

\author{Cristiane M. de Oliveira}

\author{Rhakny P.P. Araújo}
 
\affiliation{Departamento de Física, Universidade Federal de Rondônia, Campus Ji-Paraná, Ji-Paraná, Rondônia, Brazil}

\date{\today}

\begin{abstract}
Currently, the benefits of using computer simulations in the school environment are widely recognized by educators. The encouragement to use these simulations in classrooms is supported by the National Common Curricular Base, a normative document that defines essential learning for basic education. This article focuses on the exploration of the Mechanical Energy Conservation simulator, a free simulator developed with the purpose of illustrating the transformations between kinetic, gravitational potential, and elastic potential energies. This application is part of SimuFísica\textsuperscript{\textregistered}, a multilingual and multiplatform software that offers computer simulations for teaching physics, which operates both online and offline. Throughout the article, some features of the SimuFísic\textsuperscript{\textregistered} platform and the details of how the Mechanical Energy Conservation simulator works are objectively presented. Two illustrative examples are provided, demonstrating the application in action for solving mechanics problems.
\end{abstract}


\maketitle


\section{Introduction}

The advantages resulting from the incorporation of computational resources in the educational environment are widely recognized by contemporary educators. Several studies highlight the learning gains, quantitatively measured, by students who participated in lessons supported by resources such as gamification [1, 2] or computer simulations [3, 4]. These studies show improvements in motivation, engagement, visualization of abstract phenomena, and interactivity, among other aspects.

The principle of mechanical energy conservation is one of the physics topics whose teaching can be significantly enriched with the support of educational resources that go beyond traditional classes. An example of an experimental activity is a proposal involving the movement of marbles along tracks with different trajectories, aimed at increasing students' understanding of concepts related to motion and energy [5]. In parallel, in the computational domain, several theses defended in Brazil are based on didactic proposals using the Skate Park Energy simulator, part of the PhET platform [6]. The relevance of these approaches is reinforced by the National Common Curricular Base (BNCC), which highlights skills related to Specific Competence 1, concerning the knowledge of the law of energy conservation, encouraging the use of computer simulations and other applications as part of scientific literacy in both elementary and high school education [7].

In this context, we highlight in this work the Mechanical Energy Conservation simulator from the SimuFísica\textsuperscript{\textregistered} platform. Our main motivation in creating this simulator was to present to the academic community the potential of a nationally developed, free software compatible with small smartphone screens. The simulator illustrates the transformations between kinetic, gravitational potential, and elastic potential energies and can serve as support in solving problems related to the law of mechanical energy conservation. Throughout the following sections, we discuss the Mechanical Energy Conservation simulator, pointing out some details about its operation and certain limitations. We conclude the article by presenting examples of the simulator's use, involving the solution of two mechanics problems, and some final considerations.

\section{The Conservation of Mechanical Energy simulator}

The Mechanical Energy Conservation simulator is part of the SimuFísica\textsuperscript{\textregistered} platform [8, 9], a comprehensive collection of computer simulation applications developed to enhance the teaching of physics. This free-access software stands out for its versatility, being compatible with devices of different screen sizes and various operating systems. The SimuFísica\textsuperscript{\textregistered} experience can be accessed online through its website\footnote{Online access: \url{https://simufisica.com/en/}.}, or by downloading and installing it from major app stores, such as Google Play (Android)\footnote{Access via Google Play: \url{https://play.google.com/store/apps/details?id=com.simufisica}.}, App Store (iOS and iPadOS)\footnote{Access via the App Store: \href{https://apps.apple.com/br/app/simuf\%C3\%ADsica/id6449152916?l=en-GB}{https://apps.apple.com/br/}.}, Microsoft Store (Windows)\footnote{Access via the Microsoft Store: \url{https://apps.microsoft.com/detail/9N7HJVB9FMZT}.}, and Snapcraft (Linux desktop)\footnote{Access via Snapcraft: \url{https://snapcraft.io/simufisica}.}. SimuFísica\textsuperscript{\textregistered} is currently offered in three languages: Portuguese, English, and Spanish, providing wide accessibility to users of different nationalities. It is worth noting that SimuFísica\textsuperscript{\textregistered} applications are not limited to illustrating physical phenomena; they aim, more broadly, to simulate processes based on the fundamental equations of physics. Far from being mere animations, these simulations aim to provide a more faithful and dynamic representation of complex physical events.

Figure 1 shows the desktop version of the Mechanical Energy Conservation application. After pressing the ``Play'' button on the bottom toolbar, the object, symbolized by the red ball and treated as a point mass, moves along a one-dimensional trajectory consisting of up to three horizontal line segments with heights $h_1$, $h_2$, and $h_3$, identified by the dashed lines in Figure 1 and interconnected by curved sections. A mathematical expression for the trajectory’s shape, i.e., $y(x)$, can be found in Ref. [10].

\begin{figure}[htpb]
	\centering
	\includegraphics[width=0.99\linewidth]{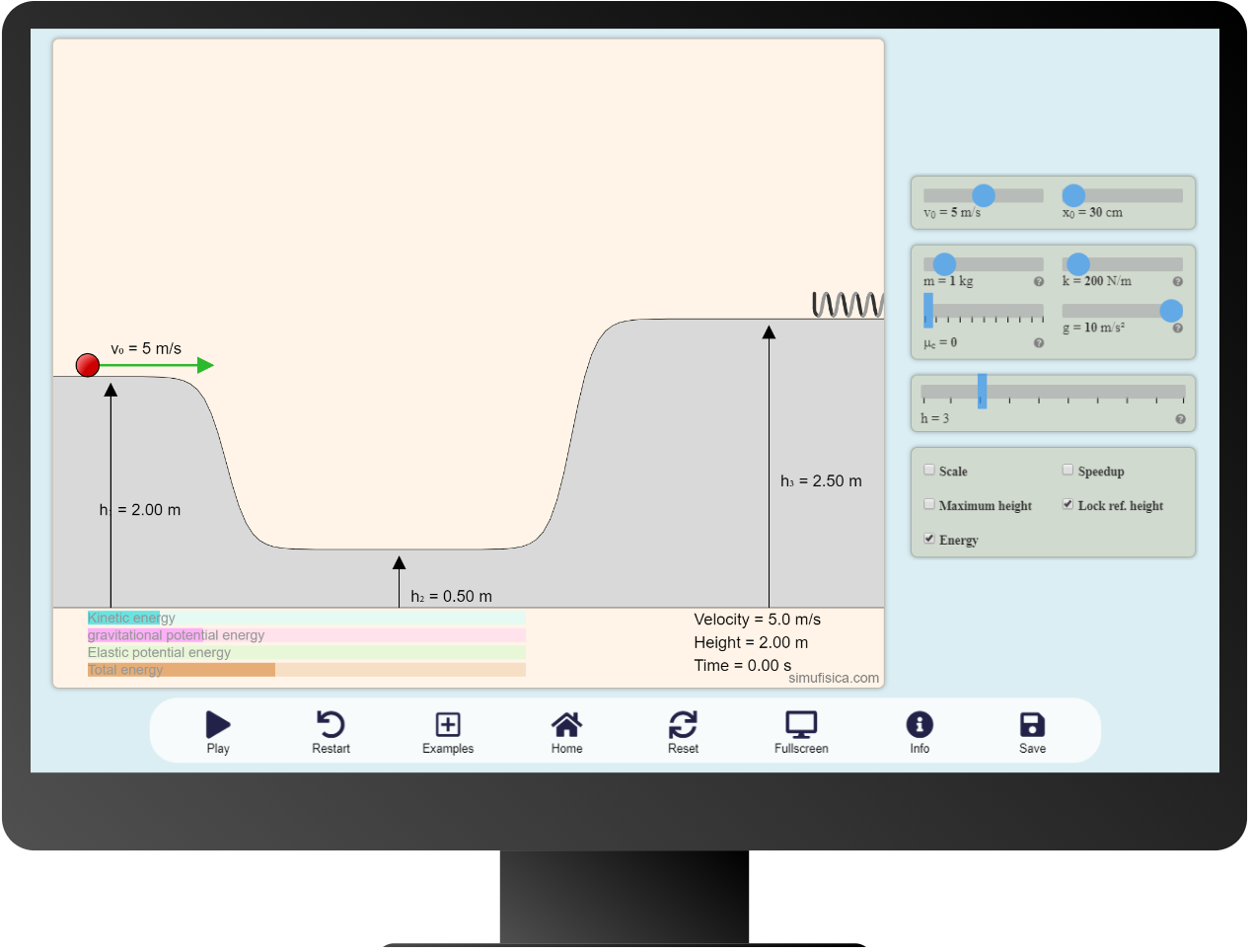}
	\caption{Conservation of Mechanical Energy application version 1.9.1 for desktops. Available at \url{https://simufisica.com/en/conservation-mechanical-energy/}.}
	\label{fig1}
\end{figure}

The simulator calculates and displays, in real time, the object's speed and height at any given moment. This operation is performed through the numerical solution of a system of nonlinear ordinary differential equations, whose derivation is available for consultation in Ref. [11]. Additionally, the application features a horizontal bar graph representing the object’s kinetic and gravitational potential energies, the spring’s elastic potential energy, and the total mechanical energy. These visual elements significantly contribute to an instant and intuitive understanding of the variables and physical principles involved in the conservation of mechanical energy.

The simulator is extremely versatile, offering a wide range of configuration options. Users can intuitively adjust the heights of the horizontal regions, either with a mouse cursor or by touch if they are using a touchscreen device. Additionally, the height reference line can be easily adjusted by unchecking the ``Lock height ref.'' option.

To provide a more personalized experience, the simulator offers additional options. The inclusion of the object's maximum height on the screen can be enabled by checking the ``Maximum height'' box. The simulation speed can be adjusted through the $h$ parameter and/or by checking the ``Accelerate'' box. Additionally, a grid with a scale can be included on the screen by selecting the ``Scale'' option. These settings allow for a precise adjustment of the simulation environment according to the user’s preferences and goals, resulting in a more engaging and personalized learning experience.

Another customization option involves the initial conditions, such as the height, which can be adjusted through both the horizontal component of the object's position ($x_0$) and the heights $h_1$, $h_2$, and $h_3$, in addition to the object’s initial scalar speed ($v_0$). Regarding the parameters, there is flexibility to modify the object's mass ($m$), the spring’s elastic constant ($k$), the coefficient of kinetic friction ($\mu_c$) in the central horizontal region (as illustrated in Fig. 2), and gravitational acceleration (g). This ability to customize the initial conditions and parameters provides users with refined control over the fundamental variables of the simulation, allowing for a broader and more specific analysis of different scenarios and their physical implications.

\begin{figure}[htpb]
	\centering
	\includegraphics[width=0.99\linewidth]{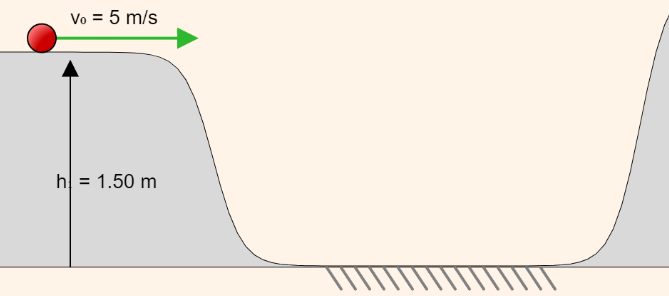}
	\caption{Part of the Conservation of Mechanical Energy app showing the flat central region with non-zero friction on the trajectory.}
	\label{fig2}
\end{figure}

The horizontal toolbar located at the bottom remains standardized across all SimuFísica\textsuperscript{\textregistered} platform simulators, although the number of buttons varies from one simulation to another. Below is a list indicating the function of each button in the Mechanical Energy Conservation simulator:

\begin{itemize}
	\item ``Play'': Starts or pauses the simulation.
	
	\item ``Restart'': Restarts the simulation while maintaining all parameters, initial conditions, and settings.
	
	\item ``Examples'' (Fig. 3): This button offers pre-configured setups, such as ``Valley with friction,'' ``Descent with spring,'' and ``Motion in a plane,'' to facilitate classroom use by teachers.
	
	\item ``Home'': Accesses the SimuFísica\textsuperscript{\textregistered} platform's homepage, where users can find other simulators.
	
	\item ``Reset'': Resets the simulation and clears all parameters, initial conditions, and settings.
	
	\item ``Fullscreen'': Switches the app to fullscreen mode.
	
	\item ``Info'': Provides the user with basic information about the simulator and an illustrative video on how it works.
	
	\item ``Save'': Saves all parameters, initial conditions, and settings in the browser (for online use) or in the app (for offline use).
\end{itemize}

\begin{figure}[htpb]
	\centering
	\includegraphics[width=0.99\linewidth]{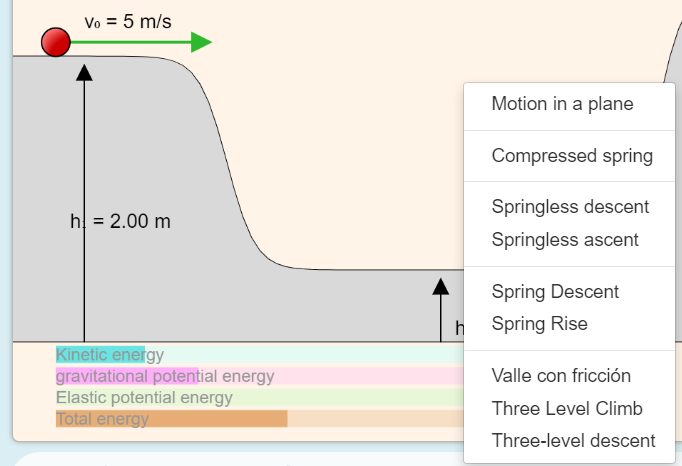}
	\caption{Part of the Conservation of Mechanical Energy app showing the available options under the “Examples” button.}
	\label{fig3}
\end{figure}

As with any program that employs numerical calculations, the Mechanical Energy Conservation application presents numerical errors that depend on the integration step used in solving the ordinary differential equations (ODEs). In this application, the solution is carried out using the fourth-order Runge-Kutta method [12]. The integration step is directly related to the $h$ parameter displayed in the application, allowing the user to configure a simulation with a higher execution speed, though this results in a corresponding increase in numerical error.

An estimate of the numerical error can be obtained by comparing the evolution of the system’s total energy (the sum of kinetic, gravitational potential, and elastic potential energies) over time with its initial value. This process is performed in a frictionless setup, where no mechanical energy dissipation occurs. The graphical result of the system's total energy, considering the default initialization configuration of the simulator (see Fig. 1) with h = 10, is presented in Fig. 4. In this case, the total mechanical energy ($E_T$) should remain constant in

\begin{equation}
	E_T = \dfrac{mv_0^2}{2} + mgh_1 = 32.5 \text{ J}.
\end{equation}

\begin{figure}[htpb]
	\centering
	\includegraphics[width=0.99\linewidth]{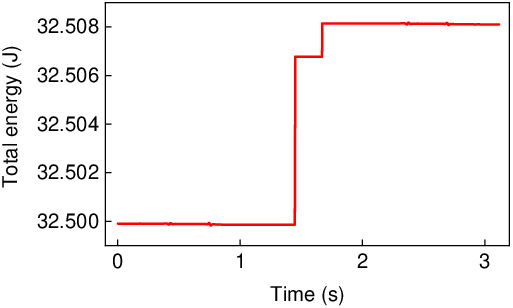}
	\caption{Total mechanical energy calculated by the Conservation of Mechanical Energy simulator over time, obtained using the console.log() function in JavaScript, inserted into the application's source code. It was calculated based on the values of speed, height, and spring deformation at all time intervals. Parameters and initial conditions considered: $v_0 = 5$ m/s, $x_0 = 30$ cm, $k = 200$ N/m, $h_1 = 2.0$ m, $h_2 = 0.5$ m, $h_3 = 2.5$ m, $m = 1$ kg, $g = 10$ m/s², $\mu_c = 0$, and $h = 10$ (numerical integration step).}
	\label{fig4}
\end{figure}

The application, however, shows a slight change in this value. The greatest variation occurs around $t = 1.45$ s, when the object begins interacting with the spring. Throughout the entire path, in which the object starts from $x_0 = 30$ cm (height = 2.0 m) and returns to the same point after colliding with the spring, there is a relative variation in the total mechanical energy value of (32.508 - 32.500) / 32.500 $\approx$ 0.025\%. This subtle discrepancy arises from the numerical error associated with this specific configuration and the value of $h$ equal to 10.

\section{Examples of using the app in problem solving}

We present here the use of the application as a support tool for visualization and calculations related to two example mechanics problems involving the principle of energy conservation.

\subsection{Finding the maximum deformation of a spring}

\textbf{Problem}: An object with a mass of 5 kg is launched with a speed of 4 m/s on a frictionless horizontal surface against a spring ($k = 500$ N/m) at rest. Find the maximum deformation of the spring.

In this and the problem in Section 3.2, we will denote the initial moment by the subscript i and the final moment by f. Since mechanical energy is conserved, it is the same at both moments:

\begin{equation}
	E_i = E_f
\end{equation}

Here, the initial moment refers to the object being launched, where the system only has kinetic energy. In the final moment, which we identify as the point when the spring reaches its maximum deformation, the system has only elastic potential energy. With this in mind, Eq. (2) becomes

\begin{equation}
	\dfrac{mv_0^2}{2} = \dfrac{kx_m^2}{2},
\end{equation}

\noindent whose solution for the maximum deformation $x_m$ is given by

\begin{equation}
	x_m = v_0\sqrt{\dfrac{m}{k}}.
\end{equation}

\noindent Inserting the numerical values, we arrive at a maximum deformation of $x_m = 40$ cm, as shown in the inset of Fig. 5.

\begin{figure}[htpb]
	\centering
	\includegraphics[width=0.99\linewidth]{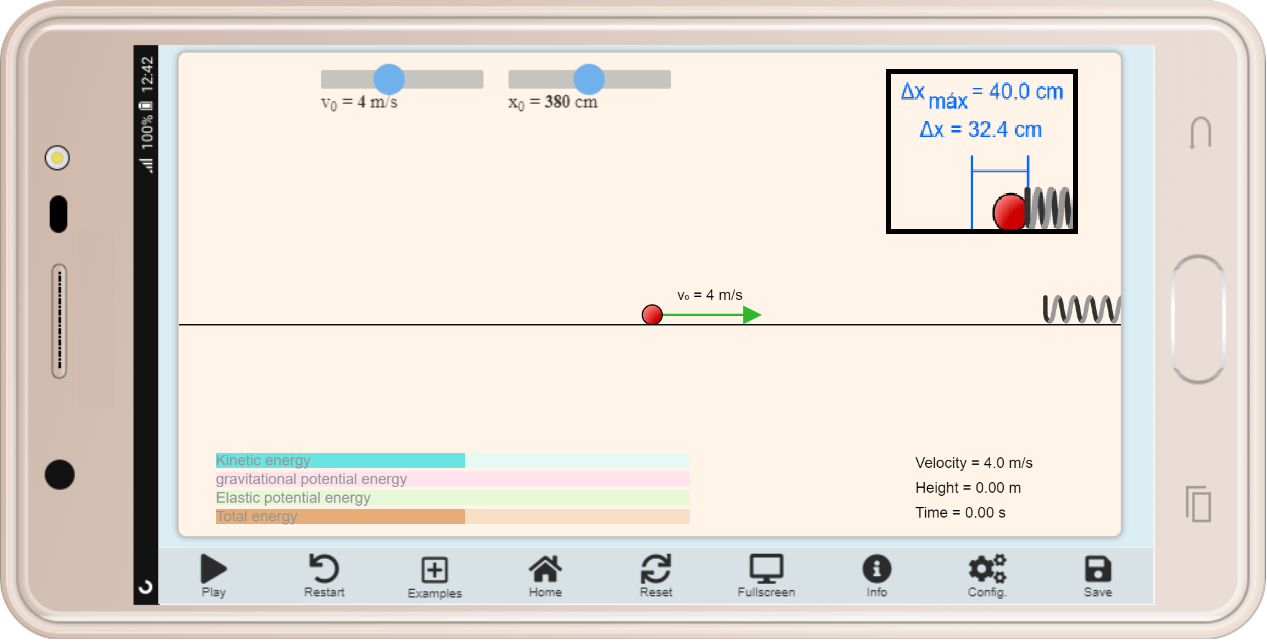}
	\caption{Conservation of Mechanical Energy simulator configured for the problem in Sec. 3.1. The inset shows the area of the application around the spring a few moments after the object reaches its maximum deformation and begins to return.}
	\label{fig5}
\end{figure}

\subsection{Motion with friction}

\textbf{Problem}: An object is released from rest at a height of 1.8 m at the edge of a valley, as shown in Fig. 6, moment (i). Assume that the lowest region, which spans 1.5 m, has a kinetic friction coefficient of 0.25. Ignore friction in other regions. What is the maximum height reached by the object after it passes through the surface with friction for the first time? How many times does the object pass through the region with friction before coming to a stop?

\begin{figure}[htpb]
	\centering
	\includegraphics[width=0.99\linewidth]{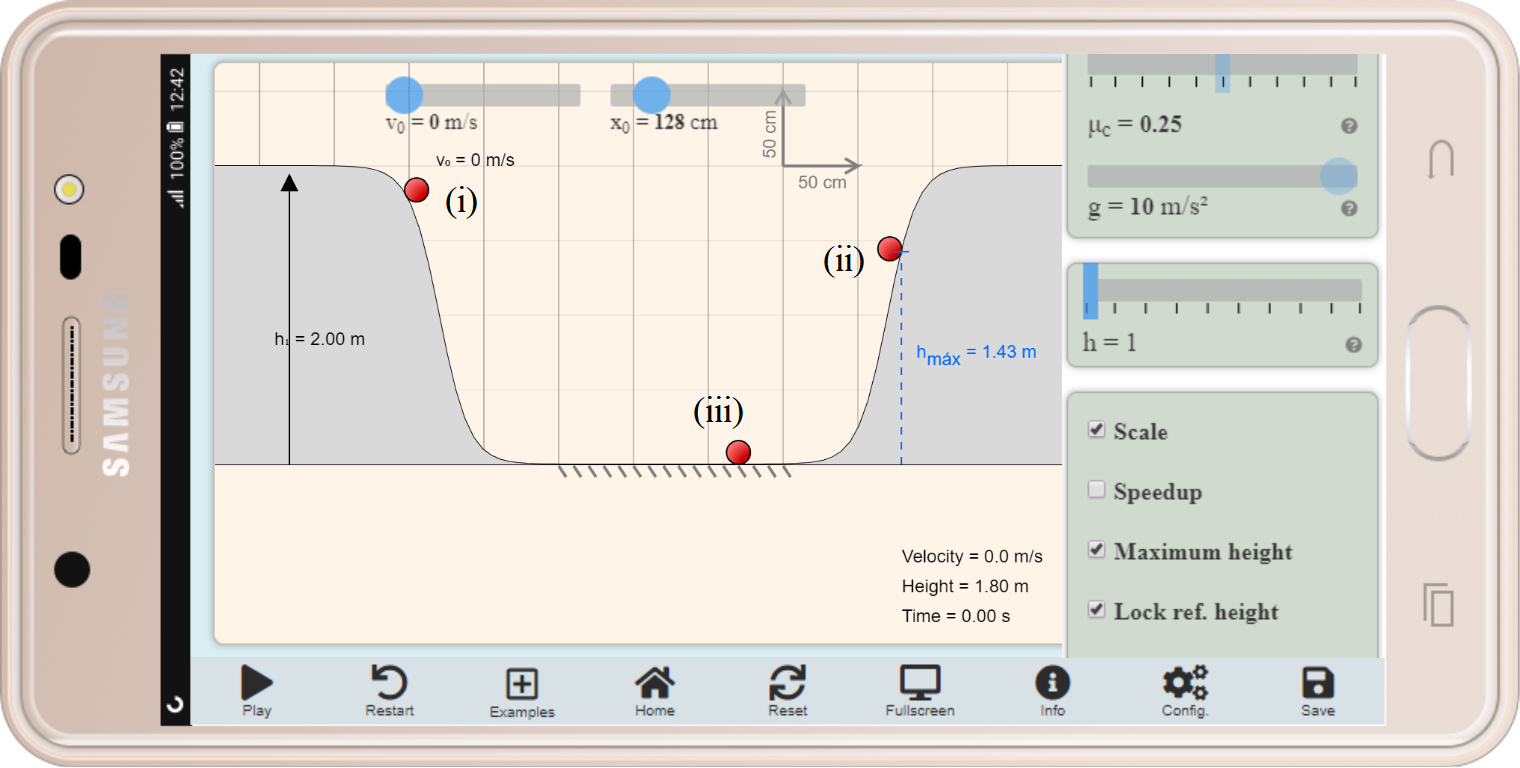}
	\caption{Conservation of Mechanical Energy simulator configured for the problem in Sec. 4.2. The image shows the superposition of three moments around the object: (i) before the start of the motion, (ii) moments after reaching the maximum height, and (iii) the moment in which the object stops in the central region.}
	\label{fig6}
\end{figure}

Since this problem involves part of the trajectory with friction, mechanical energy is not conserved. We can start by calculating the energy dissipated by friction when the object passes completely through the central region. Knowing that this region has a length of $D = 1.5$ m and that the work done by the friction force ($F_{at} = \mu_c mg$) is the product of the friction force magnitude by the distance $D$, we express the dissipated energy $E_d$ over the friction path as:

\begin{equation}
	E_d = \mu_c m g D.
\end{equation}

The gravitational potential energy of the object at its maximum height will be its initial gravitational potential energy minus the loss due to dissipation:

\begin{equation}
	mgh_f = mgh_i - \mu_c mgD.
\end{equation}

\noindent The solution for $h_f$ is given by:

\begin{equation}
	h_f = h_i - \mu_c D.
\end{equation}

Using the numerical values, we find $h_f = 1.425$ m, which answers the first part of the problem, as shown in Fig. 6. The app rounds the value, as the maximum height is always displayed with two decimal places.

To determine the number $N$ of times the object passes through the lowest part of the trajectory, we divide the total initial energy by the energy dissipated by friction during each pass, leading to the equation:

\begin{equation}
	N = \dfrac{h_i}{\mu_c D}
\end{equation}

Substituting the values, we find $N = 4.8$, indicating that the object passes through the lowest region five times before coming to a stop, with the object covering only 80\% of the path length during the last pass. This result can be seen approximately in Fig. 6, moment (iii), when comparing the object’s final position with the scale displayed by the app. Other problems solved using the app can be found in Ref. [13].

\section{Final Considerations}

This article presented the \textbf{Conservation of Mechanical Energy} simulator, part of the SimuFísica\textsuperscript{\textregistered} platform, as an educational tool for teaching physics. By exploring the growing acceptance of using computer simulations in classrooms, we observed how this application aligns with the guidelines of the National Common Core Curriculum, reinforcing its role as a valuable resource for teachers.

It is important to highlight some key features of this application. Firstly, its compatibility with mobile devices, such as tablets and smartphones, stands out. In 2022, approximately 60\% of all internet traffic involved mobile devices [14], underscoring the importance of optimizing educational simulators for these devices. Secondly, higher engagement rates are known to be achieved when applications are downloaded from app stores [14], such as Google Play and the App Store, where SimuFísica\textsuperscript{\textregistered} is available. In addition to the convenience of quick access, the ability to use the app offline is a crucial consideration, given that a lack of internet access is a reality in many schools across the country [15], or even in students' homes [16]. Additionally, in several of these schools, the adopted operating system is a Linux distribution, and SimuFísica\textsuperscript{\textregistered} is compatible, as it can be downloaded via Snapcraft, Canonical's app store.

\section*{A.I. disclosure statement}

This article was translated from Portuguese using generative AI based on the version available at \url{https://fisicanaescola.org.br/index.php/revista/article/view/173}.

\appendix

\section{Motion of a Particle in a One-Dimensional Trajectory}

	Consider the motion of a particle of mass \(m\) constrained to a given trajectory \(y=y(x)\). The particle is acted upon by the weight force \(P\) and, in certain regions of the path, by frictional force \(F_{\text{at}}\) and elastic force \(F_{\text{el}}\) from a spring, as shown in Fig. 7. Given the initial conditions \(x(0)=x_0\) and \(v(0)=v_0\), we will show how to find the position and velocity at any instant \(t>0\).
	
	\begin{figure}[htpb]
		\centering
		\includegraphics[width=0.99\linewidth]{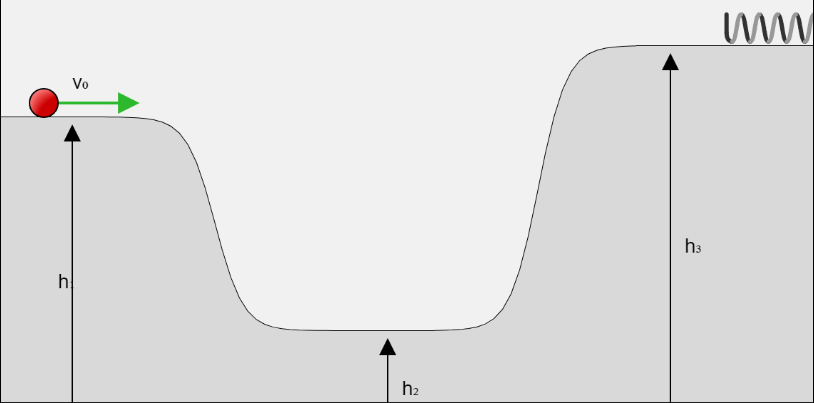}
		\caption{Motion of a particle in a one-dimensional trajectory. Based on the Mechanical Energy Conservation app.}
		\label{fig1A}
	\end{figure}

	We will consider the Lagrangian formalism of classical mechanics, which is the easiest way to arrive at the equations of motion. The kinetic energy of the particle is given by
	
	\[
	T = \frac{1}{2} m ( \dot{x}^2 + \dot{y}^2). \tag{1}
	\]
	
	Since the motion is restricted to a one-dimensional trajectory, we can eliminate the variable \(y\) by considering the constraint
	
	\[
	y = y(x). \tag{2}
	\]
	
	In this case, the kinetic energy depends only on \(x\) and \(\dot{x}\):
	
	\[
	T = \frac{1}{2} m \dot{x}^2 \left[1 + (y'(x))^2\right]. \tag{3}
	\]
	
	The potential energy \(U\) of the system has two terms: gravitational potential energy and elastic potential energy, so we write
	
	\[
	U = mg y(x) + \frac{1}{2} k (x - x_M + l_0)^2 H(x - x_M + l_0), \tag{4}
	\]
	
	\noindent where \(y(x)\) is the height of the particle relative to a given reference, \(H\) is the Heaviside step function, \(x_M\) is the position of the fixed end of the spring, and \(l_0\) is its resting length. In the app, \(l_0=70\) cm. \(H\) appears in equation (4) because the elastic force starts acting on the object only when it is in contact with the spring, that is, whenever \(x>x_M-l_0\).
	
	With equations (3) and (4), we obtain the Lagrangian of the system, still not considering the dissipation due to the work done by the frictional force:
	
\begin{align}
	L &= \frac{1}{2} m \dot{x}^2 \left[1 + (y'(x))^2\right] - mg y(x) \nonumber \\
	& \quad - \frac{1}{2} k (x - x_M + l_0)^2 H(x - x_M + l_0). \tag{5}
\end{align}

	Now we start from the Euler-Lagrange equation,
	
	\[
	\frac{\partial L}{\partial x} - \frac{d}{dt} \left( \frac{\partial L}{\partial \dot{x}} \right) = 0, \tag{6}
	\]
	
	\noindent to deduce the equations of motion. For the first term on the left side of equation (6), we can write
	
\begin{align}
	\frac{\partial L}{\partial x} &= m \dot{x}^2 y'(x) y''(x) - mg y'(x) \nonumber \\
	& \quad - k |x - x_M + l_0| H(x - x_M + l_0). \tag{7}
\end{align}

	For the second term, the calculation of the partial and total derivatives leads to
	
	\[
	\frac{d}{dt} \left( \frac{\partial L}{\partial \dot{x}} \right) = m \dot{x}^2 y'(x) y''(x) + m \ddot{x} [1 + (y'(x))^2]. \tag{8}
	\]
	
	Combining equations (7) and (8), we arrive at the following equation:
	
\begin{align}
	[1 + (y'(x))^2] \ddot{x} & + y'(x) y''(x) \dot{x}^2 \nonumber \\
	& + g y'(x) + \dfrac{k}{m} |x - x_M + l_0 | \times \nonumber \\
	& \times H(x - x_M + l_0) = 0.\tag{9}
\end{align}

	This equation has two nonlinear terms, implying that, most likely, there is no exact solution for it. We still need to include the effect of dissipation due to friction.
	
	Already anticipating a numerical solution, we will split this second-order ODE into two first-order ones. Considering that \(\dot{x} = v\) and that \(\ddot{x} = \dot{v}\), we arrive at the system of ODEs below:
	
\begin{align}
	\dot{x} & = v , \nonumber\\
	\left[1 + (y'(x))^2\right]\dot{v} & = -y'(x) y''(x) v^2 - g y'(x) - \nonumber\\
			& -\dfrac{k}{m} |x - x_M + l_0| H(x - x_M + l_0). \tag{10}
\end{align}

	The last step is to include the effect of friction. In the app, the friction coefficient \(\mu_c\) exists only in the central flat region, centered at \(x=x_a\) with width \(l_a\). We know that the kinetic friction force is given, in magnitude, by
	
	\[
	F_{\text{at}} = \mu_c mg. \tag{11}
	\]
	
	Thus, inserting \(\frac{F_{\text{at}}}{m}\) into the system (10), we finally arrive at the set of ODEs behind the Mechanical Energy Conservation simulator:
	
	\begin{align}
		\dot{x} & = v , \nonumber\\
		\left[1 + (y'(x))^2\right]\dot{v} & = -y'(x) y''(x) v^2 - g y'(x) - \nonumber\\
		& -\dfrac{k}{m} |x - x_M + l_0| H(x - x_M + l_0) \nonumber \\
		& \quad - \mu_c g \Pi\left( \dfrac{x - x_a}{l_a} \right)  \text{sign}(v), \tag{12}
	\end{align}

	\noindent where \(\Pi\) is the rectangle function, or Heaviside Pi function, defined by
	
	\[
	\Pi(x_a) =
	\begin{cases}
		0, & |x| > a \\
		1, & |x| = a \\
		\frac{1}{2}, & |x| < a.
	\end{cases} \tag{13}
	\]
	
	\noindent \(\Pi\) appears in system (12) to limit the dissipation due to friction within the interval \(x_a - \frac{l_a}{2} < x < x_a + \frac{l_a}{2}\).
	
	In the application, the system of equations (12) is solved numerically using the fourth-order Runge-Kutta method with a temporal integration step of \(0.0035h\) (in units of seconds), where \(h\) can be chosen by the user with possible values between 1 and 10. This should lead to a numerical error of around \(0.025\%\).

\end{document}